\documentstyle[prl,aps,preprint]{revtex}
\let\ssection=\section
\renewcommand{\section}{\setcounter{equation}{0}\ssection}

\newcommand{\be}{\begin{equation}}
\newcommand{\ee}{\end{equation}}
\newcommand{\ba}{\begin{eqnarray}}
\newcommand{\ea}{\end{eqnarray}}
\newcommand{\bec}{\begin{center}}
\newcommand{\eec}{\end{center}}

\def\negenspace{\kern-1.1em} 
\def\quer{\negenspace\nearrow}    

\begin{document}
\draft

%\twocolumn
\widetext
\title{{\small MZ-TH/97-18\hfill ${}$\\[1cm]}
Chiral anomaly in contorted spacetimes}

\author{Eckehard W. Mielke$^\diamond$\thanks{E-mail: ekke@xanum.uam.mx} 
and Dirk Kreimer$^\$ $\thanks{E-mail: kreimer@dipmza.physik.uni-mainz.de} \\ 
$^{\diamond}$ Departamento de F\'{\i}sica,\\
Universidad Aut\'onoma Metropolitana--Iztapalapa,\\
Apartado Postal 55-534, C.P. 09340, M\'exico, D.F., MEXICO\\
$^\$ $ Institut f\"ur Physik,\\
Universit\"at Mainz\\
55099 Mainz, GERMANY}

\maketitle
\begin{abstract}
The Dirac equation in Riemann--Cartan spacetimes with torsion 
is reconsidered. As is well-known, only the axial covector torsion  $A$, a 
one-form, couples to {\em massive} Dirac fields. Using diagrammatic 
techniques, we show that 
besides  the familiar Riemannian  term only the Pontrjagin type four--form 
$dA\wedge dA$ does arise additionally in the {\em chiral anomaly}, but {\em not} the 
Nieh--Yan term  $d\,^* A$, as has been  claimed recently. Implications for cosmic strings in 
Einstein--Cartan theory as well as for Ashtekar's canonical approach to 
quantum gravity are discussed.
\end{abstract}
%********************************************* 
\section{Introduction}
Quantum anomalies both in the Riemannian and in the
Riemann-Cartan spacetimes were calculated previously in several 
papers using different methods,
see e.g. \cite{Odintsov,Gren86,Yajima,Wies96,CZ}. However, recently 
Chandia and Zanelli \cite{ChandiaZ} 
have questioned the completeness of the earlier calculations which all 
demonstrated that the Nieh--Yan four-form \cite{NY} is irrelevant to the axial anomaly. 

For the axial anomaly, we have a couple of distinguished
features. Most prominent is its relation with the Atiyah-Singer 
index theorem \cite{At96}. But also from the viewpoint of perturbative
quantum field theory, the chiral anomaly has 
some
features which signal its conceptual importance.
There is the remarkable fact that it does not renormalize --- higher
order loop corrections do not alter its one-loop value. This very fact guarantees
that the anomaly can be given a topological interpretation.
Another feature is its finiteness: in any approach, the chiral anomaly
as a topological invariant is a finite quantity. 

Chandia and Zanelli argue that the Nieh-Yan four-form will add to this
quantity. As usual, they confront the fact that such a term,
if it is generated at all, is ill-defined, by whatever regulator
one uses. In their case, they use a Fujikawa-type approach
and propose to absorb the regulator mass in a rescaled vierbein.
We will present arguments which question the validity of such an
approach. 

%********************************************* 
\section{Gravitational Chern--Simons and Pontrjagin terms}

We are using Clifford--algebra valued exterior forms \cite{Mi87,mie86}, 
in which the constant Dirac matrices $\gamma_\alpha$ obeying 
$\gamma_\alpha\,\gamma_\beta +\gamma_\beta\gamma_\alpha=2o_{\alpha\beta}$
are saturating the index of the orthonomal coframe one--form $\vartheta^\alpha$ and its 
Hodge dual $\eta^\alpha:={}^\ast\vartheta^\alpha$ via:
\begin{equation}
\gamma:=\gamma_\alpha\vartheta^\alpha\,,\qquad
{}^\ast\gamma=\gamma^\alpha\eta_\alpha\,.\label{10-4.2}
\end{equation}

In terms of the Clifford algebra--valued {\em connection} 
$\Gamma := {i\over 4} \Gamma^{\alpha\beta}\,\sigma_{\alpha\beta}$, the 
$SL(2,C)$--covariant 
exterior derivative is given by $D=d+ \Gamma\wedge$, 
where  
${\sigma}_{\alpha\beta}= \frac{i}{2}
(\gamma_\alpha\gamma_\beta-\gamma_\beta\gamma_\alpha)$ are 
the Lorentz generators entering also in the 
Clifford-algebra valued two-form
$\sigma:={i\over 2}\gamma\wedge\gamma = {1\over 2}\,{\sigma}_{\alpha\beta}
\,\vartheta^\alpha\wedge\vartheta^\beta$.

Differentiation of these basic 
variables leads to the Clifford
algebra--valued {\em torsion} and {\em curvature} two--forms: 
\begin{equation} 
\Theta :=D\gamma =T^{\alpha}\gamma_{\alpha}\;, \qquad 
\Omega := d\Gamma +\Gamma\wedge \Gamma = 
{i\over 4}R^{\alpha\beta}\,\sigma_{\alpha\beta}\, . 
\label{tor} 
\end{equation} 

In Riemann--Cartan (RC) geometry, the {\em Chern--Simons term} for the  
Lorentz connection  reads \cite{He11}
\begin{equation} 
C_{\rm RR}  := - Tr\, \left( {\Gamma}\wedge {\Omega} - 
{1\over 3} {\Gamma}\wedge {\Gamma}\wedge  {\Gamma} \right)  \, . 
\label{CRR} 
\end{equation} 
The corresponding Pontrjagin term can be obtained 
by exterior differentiation
\be 
dC_{\rm RR}  = - Tr\, \left( {\Omega}\wedge {\Omega}\right)=
{1\over 2}\,R^{\alpha\beta} \wedge R_{\alpha\beta}\,.
\ee

Since the coframe can be regarded as the translational part of the 
Cartan connection \cite{MMNH,PRs,Lopez}, a related 
{\em translational} Chern--Simons term arises 
\begin{equation} 
C_{\rm TT}  :=  {1\over{8\ell^2}} Tr\, ( {\gamma} \wedge  {\Theta} )= 
{1\over{2\ell^2}}\, {\vartheta^\alpha}\wedge T_{\alpha}\,.  
\label{CTT} 
\end{equation} 
 By exterior differentiation we obtain the Nieh--Yan four--form \cite{NY}:
\begin{equation}
dC_{\rm TT} 
={2\over{\ell^2}} \left(T^\alpha\wedge
T_\alpha+R_{\alpha\beta}\wedge\vartheta^\alpha\wedge\vartheta^\beta\right) \,.
\label{eq:NY} 
\end{equation}

A fundamental length $\ell$ introduced here is necessary for 
dimensional reasons. This is also motivated 
by a de Sitter type \cite{GH} approach, in which the $sl(5,R)$--valued 
 connection   $\hat\Gamma =\Gamma +(1/\ell)(\vartheta^\alpha L^4{}_\alpha +  
 \vartheta_\beta L^\beta{}_4{})$ is expanded into the dimensionless linear connection 
 $\Gamma$ plus the coframe with canonical dimension $[length]$. 
The corresponding 
Pontrjagin term $\hat C_{\rm RR}$ 
splits via 
\be
\hat C_{\rm RR} =C_{\rm RR} -2 C_{\rm TT} \label{Poin} 
\ee 
into the linear one 
and the translational Chern--Simons term, see 
the footnote 31 of Ref. \cite{PRs}. (This relation has recently been
``recovered" by Chandia et al. \cite{ChandiaZ}).

%*************************************************************** 

\section{Dirac fields in Riemann--Cartan spacetime}
The Dirac Lagrangian  is given by the manifestly {\em Hermitian} 
four--form 
\begin{equation} 
 L_{\rm D}=L(\gamma,\psi,D\psi)= 
  {i\over 2}\left\{\overline{\psi}\,{^*\gamma}\wedge D\psi 
  +\overline{D\psi}\wedge{^*\gamma}\,\psi\right\}+{^* m}\, 
\overline{\psi}\psi\,, 
\label{eq:ldirac} 
\end{equation} 
for which $\overline{\psi}:=\psi^\dagger\gamma_0$ is the Dirac adjoint 
and $^* m=m\eta$ the mass term, cf. \cite{Mi87}.

The Dirac equation and its adjoint are obtained by varying $ L_{\rm D}$
independently with respect to $\overline{\psi}$ and $\psi$:
\begin{eqnarray}
  i{^*\gamma}\wedge D\psi+
  {^{*}m}\,\psi-{i\over 2}(D{^{*}\gamma})\psi
  &=& 0, \nonumber\\
  i\overline{D\psi}\wedge{^{*}\gamma}+
  {^*m}\,\overline{\psi}+{i\over 2}\overline{\psi} D
  {^{*}\gamma} &=& 0\,.  \label{Dirac}
\end{eqnarray}
If we make use of the properties of the Hodge dual and the
torsion $\Theta:= D\gamma$, the Dirac equation assumes the equivalent
form
\be
i \,^*\gamma\wedge \left(D  +{i\over 4}\,m\gamma-{1\over 2}\, T\right)\psi=0\,,  
\label{di3}
\ee
where 
\be T:= {1\over 4}\, Tr\left(\check{\gamma} \rfloor\Theta\right) =e_\alpha\rfloor T^\alpha\,, 
\qquad  A:=
  {1\over 4}\,Tr\left(\check{\gamma}\rfloor  
{^*\Theta}\right)= {1\over 4}\,^*Tr(\gamma\wedge\Theta) =   
\,^*(\vartheta^\alpha\wedge T_\alpha)\, .
\ee
are the one--forms of the trace and axial vector torsion, respectively.

Note that  torsion is also hidden in the RC covariant derivative $D$.
In order to separate out the purely Riemannian piece from torsion terms, 
we decompose the Riemann--Cartan connection $\Gamma=\Gamma^{\{\}}-K$ 
into the Riemannian (or Christoffel) connection 
$\Gamma^{\{\}}$ and the {\em contortion} one--form 
$K= {i\over 4} K^{\alpha\beta}\,\sigma_{\alpha\beta}$, obeying 
$\Theta=- [K, \gamma]$. 
Accordingly, the Dirac 
Lagrangian (\ref{eq:ldirac}) splits into a Riemannian and a 
spin--contortion piece,  cf. Refs. \cite{Mi87,mie86}: 
\ba 
L_{\rm D} &=& L(\gamma,\psi,D^{\{\}}\psi)-{i\over 2} 
\overline{\psi}\left({^*\gamma}\wedge K - K\wedge 
  {^*\gamma}\right)\psi\nonumber \\ 
&=&L(\gamma,\psi,D^{\{\}}\psi) 
  -{1\over 48}\,\left[Tr\left(\check{\gamma}\rfloor  
{^*\Theta}\right)\right] 
  \wedge\overline{\psi}\sigma\wedge \gamma\psi\nonumber\\ 
&=&L(\gamma,\psi,D^{\{\}}\psi) 
  -{1\over 4}\,A  \wedge\overline{\psi}\gamma_5\,^* \gamma\psi\,.
\label{decldirac} 
\ea 

%----------------------------------------------------------------------
\section{Classical axial anomaly}
Since $ L_{\rm D}=\overline{L}_{\rm D}=L_{\rm D}^\dagger$ is 
{\em Hermitian} as  
required, it provides us with the following {\em charge and axial current}, 
respectively, 
\be 
j= \overline{\psi}\, ^*\gamma \psi\, , \qquad 
j_5 :=
\overline{\psi}\gamma_5\,^*\gamma \psi=
{1\over 3} \overline{\psi}\sigma\wedge\gamma \psi
\, .  \label{eq:axial} 
\ee

{}From the Dirac equation (\ref{Dirac}) and its adjoint one can readily 
deduce the well--known ``classical axial anomaly" \cite{Kaku}
\be dj_5 =2imP  =2im \overline{\psi}\gamma_5\psi
\ee
for {\em massive} Dirac fields. This also holds 
in a Riemann--Cartan spacetime. If we restore chiral symmetry in the limit   
$m\rightarrow 0$, this leads to classical conservation law of the 
axial current for massless Weyl spinors, or since $dj =0$, 
equivalently, for the {\em chiral currrent} 
\be 
j_\pm :={1\over 2}  
\overline{\psi}(1 \pm\gamma_5)\,^*\gamma\psi =
\overline{\psi}_{\rm L,R}\,^*\gamma\psi_{\rm L,R}\,, \qquad dj_\pm =0\, .   
\ee

%----------------------------------------------------------------------
\section{Squared Dirac equation}
The  decomposed Lagrangian 
 (\ref{decldirac}), i.e. 
\be
L_{\rm D}=L(\gamma,\psi, D^{\{\}}\psi)
  -{1\over 4}\,A \wedge j_5 \, ,
  \label{decldirac2}
  \ee
leads to the 
following equivalent form of the Dirac equation in RC spacetime
\begin{equation} 
%i{^*\gamma}\wedge \left[D^{\{\}}+{^* m} 
%  -{1\over 12}\, A \wedge\sigma\wedge\gamma\right]\,\psi=
i \,^*\gamma\wedge \left[D^{\{\}}  +{i\over 4}\,m\gamma +
{i\over 4}\, A\gamma_5\right]\psi= 0\,.
\label{tt} 
\end{equation} 
Note that for a Riemannian covariant derivative we have $D^{\{\}}\gamma=0$.
Hence, in a Riemann--Cartan spacetime a Dirac spinor does 
only feel the {\em axial torsion} one--form $A$.

Thus the Hermitian Dirac operator in a RC spacetime is the zero form
\be
D\quer := 
i \,^* \left[\,^*\gamma\wedge \left(D^{\{\}}  +{i\over 4}\,m\gamma+
{i\over 4}\, A\gamma_5\right)\right] = 
i D\quer^{\{\}} +{1\over 4}\,\gamma_5 A\quer -m \, ,
\label{diop}
\ee
where the usual Feynman ``dagger" convention 
$\, A\quer:=\check{\gamma} \rfloor A= \gamma^\alpha e_\alpha\rfloor A= -(-1)^s
\,^* \left[\,^*\gamma\wedge A\right]$ for one--forms is used. 

For squaring this operator,  we are going to use 
the geometric identity (3.6.13) of \cite{PRs} restricted to 
Riemannian spacetime, i.e.
\be
[D^{\{\}}_\alpha\,, D^{\{\}}_\beta] = 
{i\over 4} R^{\{\}}_{\alpha\beta\mu\nu}\sigma^{\mu\nu}\,,
\ee
where $D^{\{\}}_\alpha :=e_\alpha\rfloor D^{\{\}}$ are the components of the 
Riemannian covariant derivative. 
 
Thus we obtain for the squared Dirac operator:
\ba
D\quer^2 &=& -{1\over 2} \gamma^\alpha \gamma^\beta 
\left(\{D^{\{\}}_\alpha\, , D^{\{\}}_\beta\} +
[D^{\{\}}_\alpha\, , D^{\{\}}_\beta]\right) - 
2im D\quer^{\{\}}\nonumber\\ 
 &-& {i\over 4}\gamma_5 (D\quer^{\{\}} A\quer) -{1\over 2}\gamma_5 
 \sigma^{\alpha\beta} A_\alpha D^{\{\}}_\beta +m^2 -{1\over 2}m \gamma_5 A\quer -
 {1\over 16} A\quer A\quer \nonumber\\
&\cong&-\Box -
{1\over 8}\sigma^{\alpha\beta} R^{\{\}}_{\alpha\beta\mu\nu}\sigma^{\mu\nu}
-{i\over 4}\gamma_5 (D\quer^{\{\}} A\quer) -{1\over 2}\gamma_5 
 \sigma^{\alpha\beta} A_\alpha D^{\{\}}_\beta -
  {1\over 16}A_\alpha A^\alpha - m^2\, ,
\label{sqdiop}
\ea
where 
$
\Box := \partial_\mu  \left (\sqrt{\mid g\mid } g^{\mu \nu }
 \partial_\nu \right )/\sqrt {\mid g\mid }$  is
the generally covariant Riemannian d'Alembertian operator. 
In the last step we used the Dirac 
equation. Not unexpectedly, besides the familiar Riemannian curvature  
scalar, only the axial torsion 
$A=A_\alpha \vartheta^\alpha$  contributes the the squared  
Dirac operator for {\em massive} spinor fields.

%****************************************
\section{Axial current in the Einstein--Cartan--Dirac theory}  
The Einstein--Cartan--Dirac (ECD) theory of a gravitationally 
coupled  spin 
$1/2$ fermion field provides a  {\em dynamical}  understanding 
of the axial anomaly on a classical (i.e., not quantized) level. 
 The ECD--Lagrangian reads:
\begin{equation}
L ={i\over 2\ell^2} \, Tr\left(\Omega\wedge\,^*\sigma\right)+ L_{\rm D} 
={1\over 2\ell^2}\,R^{\alpha\beta}\wedge\eta_{\alpha\beta} + L_{\rm D}\, ,
\end{equation}
where $\eta^{\alpha\beta}:={}^\ast(\vartheta^\alpha\wedge
\vartheta^\beta)$ is  dual to the unit two--form. 

The spin current of the Dirac field is given by the Hermitian three--form 
\begin{equation} 
\tau_{\alpha\beta}:={\partial L_{\rm D}\over\partial\Gamma^{\alpha\beta}} 
={1\over 8}\overline{\Psi}\left(\,^* \gamma\sigma_{\alpha\beta}+ 
\sigma_{\alpha\beta}\,^* \gamma\right)\Psi
= -\frac{1}{4}\,\eta_{\alpha\beta\gamma\delta}\,\overline{\Psi} 
\gamma_5\gamma^{\delta}\Psi\eta^{\gamma}\,, \label{eq:spingamma} 
\end{equation} 
which implies that the components 
$\tau_{\alpha\beta\gamma}=\tau_{[\alpha\beta\gamma]}$ of the spin  
current are {\em totally antisymmetric}. 

The second field equation of EC--theory, i.e. Cartan's
algebraic  relation between torsion and spin, 
\begin{equation}
-{1\over 2} \eta_{\alpha\beta\gamma}\wedge T^{\gamma}=\ell^{2} 
\tau_{\alpha\beta}, \label{carspin}
\end{equation} 
implies the following relation (cf. \cite{MMM96}) 
between the {\em axial current} $j_5$  
of the Dirac field
and the translational Chern-Simons term 
(\ref{CTT}), or, equivalent, for the  axial torsion one-form:
\begin{equation}
C_{\rm TT} \cong
\frac{1}{4}\,j_5 \,, \qquad 
A=2\ell^2\,^* C_{\rm TT}=
(\ell^2/2) \overline{\psi}\gamma_5\gamma\psi\, . 
\label{eq:shell} 
\end{equation}

Thus we find in ECD-theory 
\begin{equation}
dj_5 \cong 4 dC_{\rm TT} 
={2\over{\ell^2}} \left(T^\alpha\wedge
T_\alpha+R_{\alpha\beta}\wedge\vartheta^\alpha\wedge\vartheta^\beta\right)
\label{eq:classan} 
\end{equation}
which establishes  a link 
to the Nieh-Yan four form \cite{NY}, but only for  {\em massive} fields as will be shown 
below. If a  coupling 
to the Weyl covector $Q$ is allowed for, the term $-Q\wedge j_5$ 
occurs on the right-hand side.

This result, cf. \cite{mie86,MMM96}, holds on the level of  first 
quantization. Since the Hamiltonian of the semi-classical Dirac field is 
not bounded from below, one has to go over to second quantization, where 
the divergence of the axial current picks up anomalous terms. The question 
is whether in the vacuum expectation value $<dj_5>$ similar torsion and 
Weyl covector terms emerge, besides the usual Pontrjagin term.

However, if we restore chiral invariance for the Dirac fields 
in the limit $m\rightarrow 0$, we  find 
within the dynamical framework of ECD theory that the 
Nieh--Yan four--form tends to zero ``on shell", i.e.
\be 
dC_{\rm TT}\cong (1/4) dj_5 \rightarrow 0\, . 
\ee
This is consistent with the fact that a 
Weyl spinor does not couple to torsion at all, because  the remaining axial torsion 
$A$ becomes a {\em lightlike} covector, i.e.
$A_\alpha A^\alpha\eta =A\wedge\,^* A \cong (\ell^4/4)\,^*j_5\wedge j_5 =0$.   
Here we implicitly assume that the light-cone structure 
 of the axial covector $\,^*j_5$ is not spoiled by quantum corrections, i.e. that no 
 ``Lorentz anomaly" occurs as in $n=4k +2$ dimensions \cite{Leut}.

%********************************************************** 

\section{Chiral anomaly in quantum field theory} 
When quantum field theory (QFT) is involved, other boundary terms 
may arise in  the {\em chiral anomaly} due to the  
non--conservation of the axial current,  cf. \cite{zum,Hir,Holst}. 
An anomaly in QFT is a (classical) symmetry which is broken
by field quantization.
Such quantum violations were calculated for the chiral current 
in a torsion-free Riemannian background before \cite{torsfree}.

Now, to approach the anomaly in the context of space-time with torsion,
we will proceed by switching off the curvature and concentrate
on the last term in the decomposed Dirac Lagrangian  (\ref{decldirac2}).

Then,  this term  can be 
regarded  as an {\em external} axial covector $A$ (without 
Lorentz or ``internal" indices)  coupled to
the axial current $j_5$ of the Dirac field in an {\em initially flat} 
spacetime. By applying
the result (11--225) of Itzykson and Zuber \cite{IZ}, we find 
that only the term 
$dA\wedge dA$ arises in the  
axial anomaly, but {\em not} the Nieh--Yan type term 
$d\,^*A\sim dC_{\rm TT}$ as was 
recently claimed \cite{ChandiaZ}.
After switching on the curved spacetime of Riemannian geometry, we finally  
obtain for the axial anomaly
\be
<dj_5>= 2im <P> + {1\over 24\pi^2}
\left[Tr\left(R^{\{\}}\wedge  R^{\{\}}\right)  -
{1\over 4} dA\wedge dA\right]\,.\label{anom}
\ee
This result is based on diagrammatic techniques and 
the Pauli--Villars regularization scheme. 

In this respect, it is a typical perturbative result.
This becomes obvious if we compare it with
other perturbative results:
another option were to use a point-splitted current,
\be
j_5(x;\epsilon):=\overline{\psi}(x)\gamma_5{}^*\gamma\psi(x+\epsilon),
\ee
where $\epsilon$ is an infinitesimal four-vector in space-time.
Such an expression can be rendered invariant 
by dressing it with a path-ordered exponential
\be
\overline{\psi}(x)\gamma_5{}^*\gamma\psi(x+\epsilon)\rightarrow
\overline{\psi}(x)\gamma_5{}^*\gamma\psi(x+\epsilon){\bf P}
\exp\left\{i\int_x^{x+\epsilon} A\right\}\, .
\ee
The variation $\delta/\delta A$ of the current 
$j_5(x;\epsilon)$ is compensated by the variation of the 
exponential. As the parallel transport from $x^\mu \to x^\mu+\epsilon^\mu$
along the infinitesimal line element can be expanded perturbatively,
it is clear that the net effect of this approach is
just the standard result $<dj_5(x)>=
2im <P>-(1/96\pi^2) dA\wedge dA$ (curvature still switched off).

A further option is to use
{\em dimensional regularization}. 
If one adopts the $\gamma_5$ scheme of Ref. \cite{DK},
one can indeed immediately conclude that only the result
(\ref{anom}) can appear.
In this scheme, 
the only effect of the $\gamma_5$ problem is the replacement
of the usual trace by a non-cyclic linear functional.
The anomaly appears as the sole effect of this non-cyclicity and,
vice versa, all non-cyclic effects are related to the anomaly as it
is manifested in the triangle graphs. There is no
room for other sources of non-cyclicity apart from the very fermion
loops which produce the result (\ref{anom}).
The whole effect of non-cyclicity is to have an operator $\Delta$,
which measures the amount of violation of gauge invariance
in this scheme \cite{DK}. One has
$\Delta^2=0$, and the anomaly is in the image modulo
the kernel of $\Delta$, which summarizes the fact that in this
$\gamma_5$ scheme no other anomalous contributions are possible
beside (\ref{anom}).

So, in agreement with \cite{Odintsov,Gren86,Yajima,Wies96} 
we find no Nieh--Yan term
in the anomaly. But at this stage we have not discussed the
possibility that the torsion is not adiabatically
connected to the torsion-free case. 
Chandia and Zanelli argue that in such a case they find a
term $\sim dC_{\rm TT}$. But in their result, this term is
multiplied by a factor $M^2$. This factor $M^2$ corresponds to a
regulator mass in a Fujikawa type approach. 
This is in agreement with previous calculations, which 
obtained such ill-defined contributions
to the anomaly as well, in
other regulator methods. They were consistently absorbed
in a counterterm, and thus {\em discarded} from the final result for the
anomaly.

In contrast, Chandia and Zanelli argue that such contributions
can be maintained by absorbing the divergent factor in a rescaled
vierbein, and propose
to consider the limit $\vartheta^\alpha\to \vartheta^\alpha/M$, with 
$M\ell$ fixed. 
Here, $\ell$ is the length scale
introduced in the topological
invariant $dC_{\rm TT}$. It is part of the definition of the topological
invariant.
But then, the argument of Chandia and
Zanelli is highly problematic. There are at least three
points which seem unsatisfactory in it.

\begin{enumerate}
\item First, consider $dC_{\rm TT}$. It is, by construction, a topological
invariant (it is the difference of two Pontrjagin
classes, after all, cf. Eq. (\ref{Poin}). Now, it is actually 
{\em not} the term
$dC_{\rm TT}$ which appears as the torsion-dependent extra contribution to the
anomaly, but the term $2\left(T^\alpha\wedge
T_\alpha+R_{\alpha\beta}\wedge\vartheta^\alpha\wedge\vartheta^\beta\right)=
\ell^2 dC_{\rm TT}$. Thus, measuring its proportion in units of the
topological invariant $dC_{\rm TT}$, we find that it vanishes when we consider
the proposed limit $M\to\infty, M\ell\sim $constant. 

\item
The second point is
that we do not have to rescale the vierbein, it is consistent
to compensate the ill-defined term by a counterterm. 
This implies
that consistently a renormalization condition can be imposed which
guarantees that the anomaly has the 
value (\ref{anom}).
Even if Chandia and Zanelli render their extra term finite by a rescaling,
they have to confront the fact that a 
(finite) renormalization condition can be imposed
which settles the anomaly at this value.
\item
Finally, it is well-known that usually the appearance of a chiral
anomaly is deeply connected with the presence of a conformal anomaly
\cite{Salam,Ellis,Kae}. This makes sense: usually,
conformal invariance is lost due to the (dynamical) generation of a scale.
But this is the very mechanism which destroys chiral invariance as well.
Thus, one would expect the Chandia and Zanelli argument to fail,
as it tries to combine strict conformal
invariance with a chiral anomaly. 

\end{enumerate}

Summarizing, our conclusions 
deviate sharply from the interpretation Chandi and Zanelli
propose.

In the limit $m\rightarrow 0$, we obtain for the  
chiral anomaly Eq. (\ref{anom}).
Depending on the asymptotic helicity states, there occur  
contributions of topological origin of the {\em Riemannian} 
Pointrjagin or Euler term, 
respectively. Interesting enough, there is a Pointrjagin type contribution   
$dA\wedge dA$ from axial torsion in Riemann--Cartan spacetime.  
Its role for the topology of contorted spacetimes is rather unexplored, cf.
\cite{Regge,DKZ}.

In the next two sections we will further strengthen our argument.
We will consider a situation where a cosmic string with 
a torsion line defect is present. This is a situation where torsion
is indeed realized in a discontinous manner.
Nevertheless, we will see that the Nieh-Yan term vanishes identically.
Also, we will discuss the heat kernel method in some detail,
pointing out the various scaling properties which are related with
the fact that $K_2$ has the wrong dimensionality in four dimensions.

%***************************************************
\section{Spinning Cosmic string with torsion}
In order to analyze also spacetimes with {\em torsion defects} 
or singularities, 
let us consider a cosmic string solution within 
the Ein\-stein--Cartan (EC) theory governed by the field equations  
\begin{equation}
-{1\over 2}\,
\eta_{{\alpha}{\beta}{\gamma}}\wedge R^{{\beta}{\gamma}}=
\ell^{2}\,\Sigma_{\alpha}\,,  \label{ECeq}
\end{equation}
and the Cartan relation (\ref{carspin}).

Let us adopt here the convention that  $x^\alpha$ together with $y^\alpha$ are 
spacelike orthogonal vectors which span the $(x, y)$--plane orthogonal to 
the $(t,z)$--plane, the world sheet of the string. The 
corresponding one--forms are denoted by capital letters, i.e.
\begin{equation}
X:=x_\alpha\, \vartheta^\alpha \, ,\qquad 
Y:=y_\alpha\, \vartheta^\alpha \, . \label{VEC}
\end{equation}
Moreover, the  vector $n^\alpha$ is a timelike unit vector normal 
to the hypersurface
with $n^\alpha\, n_\alpha=s$, the signature $s$ of spacetime. 

Following Soleng \cite{SO}, cf. Anandan \cite{A94},
we assume that 
three--forms $\Sigma_{\alpha}$ and $\tau_{{\alpha}{\beta}}$ 
of the energy--momentum and spin current, repectively, vanish  outside of  the string, 
whereas  ``inside" they
are 
{\em constant}, i.e.
\begin{equation}
 \Sigma_{\alpha}=\varepsilon\,\vartheta_\alpha\wedge X\wedge Y\, , \qquad
 \tau_{\alpha\beta}= \sigma\eta_{\alpha\beta\gamma}\, n^\gamma
\wedge X\wedge Y\, .
\end{equation}
The constant parameters $\varepsilon$ and $\sigma$ of this 
{\em spinning string} are related to the exterior vacuum solution 
by appropriate matching conditions. 
For the related solution with {\em conical singularities} and torsion of 
Tod \cite{TO},   $\varepsilon$ and $\sigma$ are {\em delta distributions} 
at the location of the string. From the specification (\ref{VEC}) of the 
one--forms $X$ and $Y$ it 
can easily be infered that the only nonzero components are 
$\Sigma_{\hat 0}\neq 0$, $\Sigma_{\hat 3}\neq 0$ and 
$\tau_{\hat 1\hat 2} =-\tau_{\hat 2 \hat 1} \neq 0$.  

Since $x^\alpha \Sigma_{\alpha}= y^\alpha \Sigma_{\alpha}=0$, 
contractions of the first EC field equation (\ref{ECeq}) 
with $x^\alpha$ and $y^\alpha$ 
reveal that
$x^{[\alpha}y^{\beta]}\, R_{\alpha\beta}=R_{\hat 1\hat 2}
=-R_{\hat 2\hat 1}\neq 0$ are the only nonvanishing components of the RC 
curvature. 
{}From its covariant
expression \cite{Ma95} 
\begin{equation}
R^{{\alpha}{\beta}}
=\varepsilon\ell^2\, x^{[\alpha}y^{\beta]}\, X\wedge Y \label{COCU}
\end{equation}
there follows  the identity  
\begin{equation}
R_\beta{}^{\alpha}\wedge \vartheta^\beta=
{\varepsilon\ell^2\over 2}
 (x^\alpha\, Y\wedge X\wedge Y-
y^\alpha\, X\wedge X\wedge Y  )=0\, .  \label{Rid}
\end{equation}

{}From the Cartan relation (\ref{carspin}) we find for the torsion 
\begin{equation}
 T^{\alpha}=-2 \sigma\ell^2\, n^{\alpha}\, X\wedge Y 
\quad\Rightarrow\quad 
T_\alpha\wedge T^{\alpha}=4s\sigma^2\ell^4\, X\wedge Y\wedge X\wedge Y=0 \,.
\label{Torid}
\end{equation}
 
Recalling that $N^\alpha= n\rfloor \vartheta^\alpha$ is 
the lapse and shift vector in the (3+1)--decomposition 
of the ADM formalism, 
the corresponding coframe and connection can now explicitly be obtained 
after applying a finite boost to the usual conical metric of the 
cosmic string, cf.\cite{A94}
\begin{eqnarray}
\vartheta^{\hat 0} &=& dt + \ell^2\sigma \rho^{\ast 2} 
         [1-\cos(\rho/{\rho^{\ast}})  ] d\phi \nonumber \\
\vartheta^{\hat 1} &=& d\rho\,, \quad\qquad
\vartheta^{\hat 2} = \rho^{\ast} \sin(\rho/{\rho^{\ast}}) d\phi\,, \quad\qquad
\vartheta^{\hat 3} = dz \, , \nonumber \\  
\Gamma^{{\hat 1}{\hat 2}} &=&  \cos(\rho/{\rho^{\ast}}) d\phi = -
\Gamma^{{\hat 2}{\hat 1}} \, .
\label{EXCO}    
\end{eqnarray}

Thus from the identities (\ref{Torid}) and (\ref{Rid}), we can infer that 
the Nieh--Yan term (\ref{eq:NY}) {\em vanishes identically} for this 
example of a  spinning cosmic 
 string exhibiting a {\em torsion line defect}. The same holds for the 
 stationary cosmic string solution of 
Letelier \cite{Le95}, see also Ref. \cite{A96}, where the vector normal 
to the hypersurface is generalized to  
$n^\alpha=(1,0,0, \tau/\sigma)$.

%***************************************************

\section{Comparison with the heat kernel method} 
In the heat kernel approach, there exists for small $t\rightarrow +0$ 
an asymptotic expansion of the 
kernel in $n$ dimensions:
\begin{equation}
K(t,x,x,D\quer^2)=(4\pi)^{-n/2}\sum_{k=0}^\infty t^{(k-n)/2} 
K_k(x,D\quer^2)\,. \label{exp}
\end{equation}
The coefficients $K_k(x,{\cal D}^2)$, $k=0,1,\dots$ 
are completely determined by the form of the 
second-order differential operator $D\quer^2$, which is positive for 
 Euclidean signature ${\rm diag}\; o_{\alpha\beta}=(-1, \cdots, -1)$. For 
 odd $k=1,3,\dots$ these
coefficients are zero, while the first nontrivial terms 
\cite{Yuri1,Yuri4,Yajima}, which potentially 
could  contribute to the axial anomaly,  read 
\ba 
 Tr(\gamma_5 K_2) &=& -d\,^* A\, ,  \nonumber\\
 Tr(\gamma_5 K_4) &=& {1\over 6} 
\left[Tr\left(R^{\{\}}\wedge  R^{\{\}}\right) -
{1\over 4} dA\wedge dA +d{\cal K}\right]\,.
 \ea

However, there is an essential difference in the physical dimensionality of the 
terms $K_2$ and 
$K_4$. Whereas in $n=4$ dimensions the Pontrjagin type term $K_4$ is 
dimensionless and thus multiplied by $t^{(k-4)/2}=1$,  the term 
$K_2\sim d\,^* A = 2\ell^2 dC_{\rm TT}$ carries dimensions. 
Since a {\em massive} Dirac spinor has canonical dimension $[length]^{-3/2}$, it scales as 
$\psi \sim m^{3/2}$. Moreover, in Fujikawa method, cf. \cite{Kaku}, the term 
$t=1/M^2$ is related to the regulator mass $M\rightarrow \infty$. Then the second order 
term in the heat kernel expansion scales as 
\be
-{1\over t} K_2={2\ell^2\over t}dC_{\rm TT}\cong {\ell^2\over 2t}dj_5 =
{im\ell^2\over t}\overline{\psi}\gamma_5\psi
\sim 
\ell^2 M^2 m^4\rightarrow 0\,. 
\ee
If we assume in the renormalization procedure, that the 
fundamental length $\ell$ 
 does not scale (no running coupling constant), the second order term in the  
 heat kernel expansion will tend to zero in 
the limit $m\rightarrow 0$.
In the case $m\neq 0$, this 
term diverges and the Fujikawa regulator method cannot be applied. To rescale 
the coframe by $\vartheta^\alpha \rightarrow  \vartheta^\alpha/\ell$ does not 
help, since this would change also the dimension of the Dirac field, in 
order to retain  the physical dimension $[\hbar]$ of the Dirac action.

 Then our conclusion is that that Nieh--Yan term $dC_{\rm TT}$ does NOT 
contribute to the {\em chiral anomaly} in $n=4$ dimensions, neither classically nor in quantum 
field theory, as pointed out before.
This is in sharp contrast to the recent paper of Chandia 
\cite{ChandiaZ} et al. 
We once more stress the interrelation between the scale and chiral invariance
\cite{Salam,Ellis,Kae}.
Since renormalization amounts to a continous scale deformation, 
only the Riemannian part of the Pontrjagin term contributes 
to the topology of the chiral anomaly, cf. \cite{WZ,Mi85}.  On would surmize that 
in  $n=2$ dimensional models only the term $d\,^* A$ survives in the heat kernel expansion, since it then has the correct   
dimensions. However, it is well-known \cite{PRs} that in 2D the axial 
torsion $A$
vanishes identically.
 
%The role of spinors for the index theorem and in the $4D$ Donaldson invariants
%via Seiberg--Witten equation has recently been reviewed by Atiyah
%\cite{At96}.

\section{Discussion}
We have shown that the translational Chern--Simons boundary term 
$dC_{\rm TT}$,
which  in the canonical formulation \'a la Ashtekar plays the role of  
a generating functional \cite{Mi92} for 
chiral (self-- or antiself-dual) variables in Einstein--Cartan--Dirac theory as well 
as in simple supergravity \cite{mie86}, is not affected by the 
chiral anomaly. The appearance of the Riemannian 
Pontrjagin term $dC_{\rm RR}$ could pose some problems for 
the canonical approach of gravity \'a la Ashtekar. This is likely the case,  
although the state 
\be
\Psi_\Lambda =\exp\left({3\over\Lambda} 
\int_{M_3}\underline{C}_{\rm RR}\right) 
\ee
involving
the tangential complexified Chern--Simons term $\underline{C}_{\rm RR}$ is known to solve 
the Hamiltonian constraint of gravity with cosmological constant $\Lambda$ 
in the loop approach \cite{Br92,Gr96}. Moreover, the 
additional Pointrjagin type term $d(A\wedge dA)$ 
arising from  
the axial torsion $A$, necessarily present in the case gravitationaly 
coupled Dirac and Rarita--Schwinger fields, 
could turn out to be a major obstacle 
for completing 
the canonical approach of gravity a la Ashtekar, on top of 
the open issue of reality conditions, cf. \cite{Thie96}.

\acknowledgments
We would like to thank Alfredo Mac\'{\i}as, Hugo Maroles--T\'ecotl, and 
Yuri Obukhov
for useful hints and comments on a preliminary version.
This work was partially supported by  CONACyT, grant No. 3544--E9311,
and by the joint German--Mexican project KFA--Conacyt E130--2924. 
One of us (E.W.M.) acknowledges the support by the short--term
fellowship 961 616 015 6 of the German Academic Exchange Service (DAAD), 
Bonn. D.K.~acknowledges support by a Heisenberg fellowship of the DFG.

%\begin{thebibliography}{999}%for Latex 
 
\end{document}